\documentclass[aps,prl,amssymb,superscriptaddress,nofootinbib,twocolumn]{revtex4}
\usepackage[paperwidth=612.0pt, paperheight=792.0pt]{geometry}
\usepackage{amsmath,amsfonts,amsthm,amssymb}
\usepackage{braket}
\usepackage{setspace}
\usepackage{amsmath}
\usepackage{appendix}
\usepackage[ansinew]{inputenc}
\usepackage{bbm}
\usepackage{bm}
\usepackage{amsbsy}
\usepackage{dsfont} % for symbols like the identity: \mathds{1}
\usepackage{graphicx} % for graphics
\usepackage{epsfig}
\usepackage{epstopdf}
\usepackage{dsfont}
\usepackage{color}
\usepackage[colorlinks]{hyperref}
\usepackage[figure,table]{hypcap}
\usepackage{enumerate}%allows different styles of enumerate environment

\begin{document}

\title{Experimental access to higher-dimensional entangled quantum systems using integrated optics}

\author{Christoph Schaeff}
\thanks{Corresponding author}
\email{christoph.schaeff@univie.ac.at}
\affiliation{Faculty of Physics, University of Vienna, Boltzmanngasse 5, 1090 Vienna, Austria}
\author{Robert Polster}
\affiliation{Faculty of Physics, University of Vienna, Boltzmanngasse 5, 1090 Vienna, Austria}

\author{Marcus Huber}
\affiliation{Universitat Autonoma de Barcelona, 08193 Bellaterra, Barcelona, Spain}
\affiliation{ICFO-Institut de Ciencies Fotoniques, 08860 Castelldefels, Barcelona, Spain}

\author{Sven Ramelow}
\affiliation{Faculty of Physics, University of Vienna, Boltzmanngasse 5, 1090 Vienna, Austria}

\author{Anton Zeilinger}
\thanks{Corresponding author}
\email{anton.zeilinger@univie.ac.at}
\affiliation{Faculty of Physics, University of Vienna, Boltzmanngasse 5, 1090 Vienna, Austria}
\affiliation{Institute for Quantum Optics and Quantum Information (IQOQI), Austrian Academy of Sciences, Boltzmanngasse 3, 1090 Vienna, Austria}
%\affil[4]{170 Clark Hall, School of Applied and Engineering Physics, Cornell University, 14853 Ithaca NY, USA}

\begin{abstract}
Integrated optics allow the generation and control of increasingly complex photonic states on chip based architectures. Here, we implement two entangled qutrits $\textendash$ a 9-dimensional quantum system $\textendash$ and demonstrate an exceptionally high degree of experimental control. The approach which is conceptually different to common bulk optical implementations is heavily based on methods of integrated in-fiber and on-chip technologies and further motivated by methods commonly used in today's telecommunication industry. The system is composed of an in-fiber source creating entangled qutrit states of any amplitude and phase and an on-chip integrated general Multiport enabling the realization of any desired local unitary transformation within the two qutrit 9-dimensional Hilbert space. The complete design is readily extendible towards higher-dimensions with moderate increase in complexity. Ultimately, our scheme allows for complete on-chip integration. We demonstrate the flexibility and generality of our system by realizing a complete characterization of the two qutrit space of higher-order Einstein-Podolsky-Rosen correlations.
\end{abstract}

%\setboolean{displaycopyright}{true}

\maketitle
%\thispagestyle{fancy}
%\ifthenelse{\boolean{shortarticle}}{\abscontent}{}

\section{Introduction}

Higher-dimensional entangled quantum states are highly relevant for fundamental questions in quantum physics and have gained increasing practical relevance due to their distinguishing properties compared to qubit states \cite{Ref1,Ref2,Ref3,Ref4,Ref5}. Moreover, there is a wide range of open theoretical questions for quantum systems not composed of multiple qubits \cite{Ref6}. The reasons for the difficulty in implementing higher-dimensional systems are mainly technical in nature. In particular, when path encoded photonic states are used phase stability quickly becomes very challenging using a bulk optical approach, traditionally used in many quantum optical experiments. Consequently, bulk optical designs are restricted in their complexity thereby hindering progress towards higher-dimensional quantum systems. 
Here we choose a conceptually different experimental approach. We utilize recent progress within the area of optical telecommunication in management and automatization of high numbers of complex optical devices. In particular, we aim at combining state-of-the art optical fiber network and integrated photonic circuit technology with methods of quantum optics.  The latter has already proven successful in different quantum experiments utilizing various integrated photonic devices \cite{Ref7,Ref8,Ref9,Ref9b,Ref10,new1,new2}. Eventually, an integrated optics approach allows for complete access to complex quantum systems with a high degree of flexibility and automatization, good scalability towards higher-dimensional realizations, as well as direct control over the parameters defining the quantum system. Reconfigurable integrated quantum circuits for qubits have been shown so far in a silica-on-silicon architecture \cite{Ref9b}.
We are not only going significantly beyond this work by demonstrating universal unitary operations in two local $9$-dimensional Hilbert spaces, but thereby also show the scalabilty of our approach towards even higher-dimensional states. Moreover, we achieve interfacing our universal multi-port chips with a maximally path-entangled qutrit source - something that we demonstrate here for the first time and show how the source-design, phase-stabilization and interfacing scales very favorable for extending it to higher dimensions. Lastly, we enter a new wavelength regime for complex integrated quantum circuits, by operating in the telecommunication band around $1550$ nm. This makes our approach directly compatible to fiber-optic networks and quantum communication applications and is technologically highly promising, as this wavelength regime features extremely mature fiber technology and is the main focus of technological advancements in integrated optics.
In the context of higher-dimensional quantum systems, higher-order non-local correlations between two (entangled) particles are of importance as they represent a fundamental distinguishing property to the classical world \cite{Ref11}. However, controlled creation and transformation in addition to $N$ outcome projective measurements of the higher-dimensional entangled system are required. 
All these requirements are met by our implemented system. We will present the concepts of our approach along with a first realization of a fully controlled and automated entangled two qutrit $9$-dimensional quantum system. The final setup allows us to completely map the two-qutrit correlation space and to compare it to theoretic predictions.

\section{Theory}
\label{sec:theory}
\begin{figure}[htbp]
\centering
\fbox{\includegraphics[width=\linewidth]{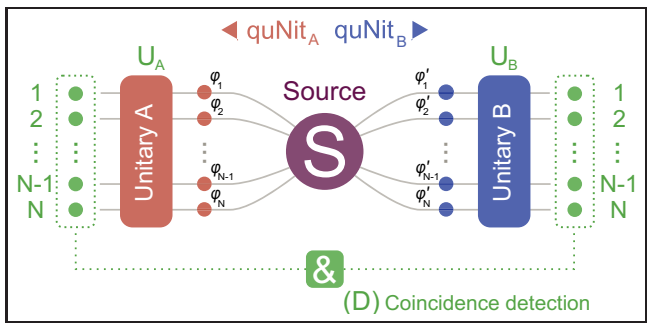}}
\caption{Concept of an entangled two quNit $N^2$-dimensional quantum experiment. A source (S) creates the path-encoded entangled two quNits state. After separation each quNit passes $N-1$ phase shifters followed by a local unitary transformation (U). Eventually the state is detected and coincidence detection events are recorded (D).}
\label{fig:theory}
\end{figure}
A general quantum-optical configuration  is illustrated in Figure \ref{fig:theory}. For $N=2$ it represents the classic text book case of a basic two qubit ($N=2$) quantum system. Generally the system is composed of three components: the creation (S), transformation (U) and detection (D) of a quantum state. A source (S) creates two particles each composed of N modes, a so-called quNit, forming the entangled state
\begin{equation}
|\psi_N\rangle=\sum_{k=1}^{N}A_ke^{\phi_k-\phi_k'}|kk\rangle\,.
\end{equation}
After separation, each quNit A and B is transformed by a local unitary transformation (U):
\begin{equation}
|\psi_N'\rangle\rightarrow(U_A\otimes U_B)|\psi_N\rangle\,.
\end{equation}
The experiment concludes with the measurement e.g. single photon detection of both quNits A and B corresponding to an underlying projective measurement
\begin{equation}
P_{ab}=|a\rangle\langle a|\otimes|b\rangle\langle b|\,,
\end{equation}
 of the quantum state $|\psi\rangle$ . The expectation value $\langle P_{ab}\rangle$ relates to the number of detected photon pairs and represents the measurement result.\\
Although, for $N=2$ Figure \ref{fig:theory} becomes a conceptually simple example it has revealed a number of quantum effects with consequences ranging from theoretical insights to prospects of applications. Quantum entanglement, manifesting itself in so-called Einstein-Podolsky-Rosen (EPR) correlations, is the first example to name \cite{Ref12}. Interestingly, although different sets of EPR correlations exist in any dimension $N$, only for $N=2$ does the number of different perfect correlations coincide with the number of all theoretically possible correlations \cite{Ref12}. Naturally, the question arises of what is to come for $N>2$. In contrast to a qubit system, a single qutrit ($N=3$) cannot be explained by non-contextual approaches \cite{Ref13}. For $N\geq4$  infinite classes of symmetric beam splitters can exist with consequences for the observation of higher-order entanglement \cite{Ref12}. At $N=6$ the question of the existence of a complete set of mutually unbiased basis is still unanswered \cite{Ref14}. Our motivation is to obtain experimental access to such quantum systems providing the basis for future tests and verifications of theoretical concepts. \\
It is worthwhile mentioning a particular experiment can be considered as one element within the full range of possible experiments covered by the general configuration in Figure \ref{fig:theory}. In contrast, we do not aim towards a single predefined measurement but rather towards a general platform allowing for a great range and flexibility within the space of all possible experiments covered by the configuration in Figure \ref{fig:theory}.

\section{Experimental Implementation}
In principle, a photonic quNit system can be encoded using any degree of freedom. However, physical and technical constraints pose a limit. Here, path-encoding will be utilized. In some way it may be considered as the most direct way of encoding quantum information into a system. Of immediate advantage is the possibility of near perfect (experimental) separability of path-encoded modes via spatial separation. Additionally, there is no fundamental limit on the number of paths. Choosing path-encoding as the starting point methods of on-chip and fiber optics offer exceptional mode and path control. The system is built using only integrated components and can be deconstructed into three parts: Source (S), transformation (U) and detection (D) as illustrated in figure~\ref{fig:theory}.
\subsection{Source of entangled qutrits (S)}
We have previously presented a scalable in-fiber source of path-encoded entangled quNits at telecom wavelengths \cite{Ref15}. It uses N non-linear crystal waveguides to create photon pairs of same polarization but different wavelength centered around 1550nm. All crystals are coherently pumped by a common 775nm pump beam, split by a sequence of beam splitters. The result is a superposition of pair creation events occurring at one of the N crystals. This is followed by 100GHz in-fiber filters separating and filtering each photon pair by wavelength (1551.7nm, 1548.5nm). A number of optical in-fiber devices ensure the matching of optical properties (spectrum, coherence, arrival time, loss, polarization), making all pair creation events indistinguishable from each other. Thereby an entangled two quNit state is obtained.
The main features of this design are its high brightness and quality of the entangled state, high robustness due to a fully fiber-integrated approach, compatibility to standard path-encoded integrated optical circuits along with a low propagation loss and good scalability in complexity when increasing the dimension of the system. We have extended the source of two entangled qubits following the scheme previously presented \cite{Ref15} to two entangled qutrits ($N=3$). 
\subsection{The Multiport (U)}
The so-called general Multiport (MP) allows an experimenter to realize any unitary transformation defined by $N^2-1$ parameters. Here it is realized using a combination of phase shifters and beam splitters. The setting of each optical device directly relates to a parameter of the unitary transformation. Depending on the phase and reflectivity settings any unitary transformation is realized \cite{Ref12}, \cite{Ref16}. In order to effectively manage the internal complexity of the device we have chosen an integrated photonic implementation.

Light-guiding Silica waveguides of low index contrast ($\Delta \approx 1.5\%$) optimized for C-band operation (cut-off wavelength $\lambda_{cutoff} \approx 1450nm$) are used. Waveguide modes closely match to the mode of a single mode fiber at $1550$nm. Beam splitters are realized by evanescent coupling of two waveguides brought to proximity. Two $50/50$ beam splitters form a Mach-Zehnder interferometer with phase shifters in each arm representing a tunable beam splitter of arbitrary reflectivity. Phase shifters are realized via the thermo-optical effect using heating electrodes above the photonic circuit on top of the cladding layer. Trenches are etched around phase shifters to ensure high thermal insulation. All phase shifters are permanently connected using a custom electronic probe card mounted from the top onto the chip. The $3\times3$ MP circuit features $6$ beam splitters and $21$ phase shifters (not all theoretically required). The chip is thermally stabilized ensuring optimal operating temperature. Total insertion loss including in- and out-coupling amounts to approximately $9$dB. The length of the $3\times3$ MP circuit is $6.5$cm at a waveguide spacing of $250\mu$m. The device is permanently packaged to standard optical fibers and electric connectors serving as a convenient optical and electrical interface. The complete housing has a size of less than $10cm\times5cm\times5cm$. Any unitary transformation entered by the user is automatically realized by a computer program translating the entered unitary matrix down to independent heating currents. Switching the Multiport to a unitary transformations requires $t<100$ms. However, we expect the actual switching time to be significantly lower. Comparable Silica circuits based on the thermos-optical effect have reached switching times of on the order of $1$ms \cite{Ref15b}. For a single Silica switch switching times on the order of $50\mu$s have been reported \cite{Ref15c}. Nonetheless, the current switching time is negligible compared to common single photon detection integration times of on the order of $10$ seconds. Therefore, switching the Multiport does currently not limit the experimental procedures in any way. We have thus not further focused on the issue. Due to the high level of control and automation the MP can be considered a black-box device. Any unitary transformation entered by the experimenter is automatically realized.
 %Its optical implementation has been previously demonstrated in detail [to be published].
%The device is permanently packaged to standard optical fibers and electric connectors serving as a convenient optical and electrical interface. Automated control mechanisms manage all internal parameters meaning the device can be treated as a black-box.
\begin{figure*}[t]
\centering
\fbox{\includegraphics[width=\linewidth]{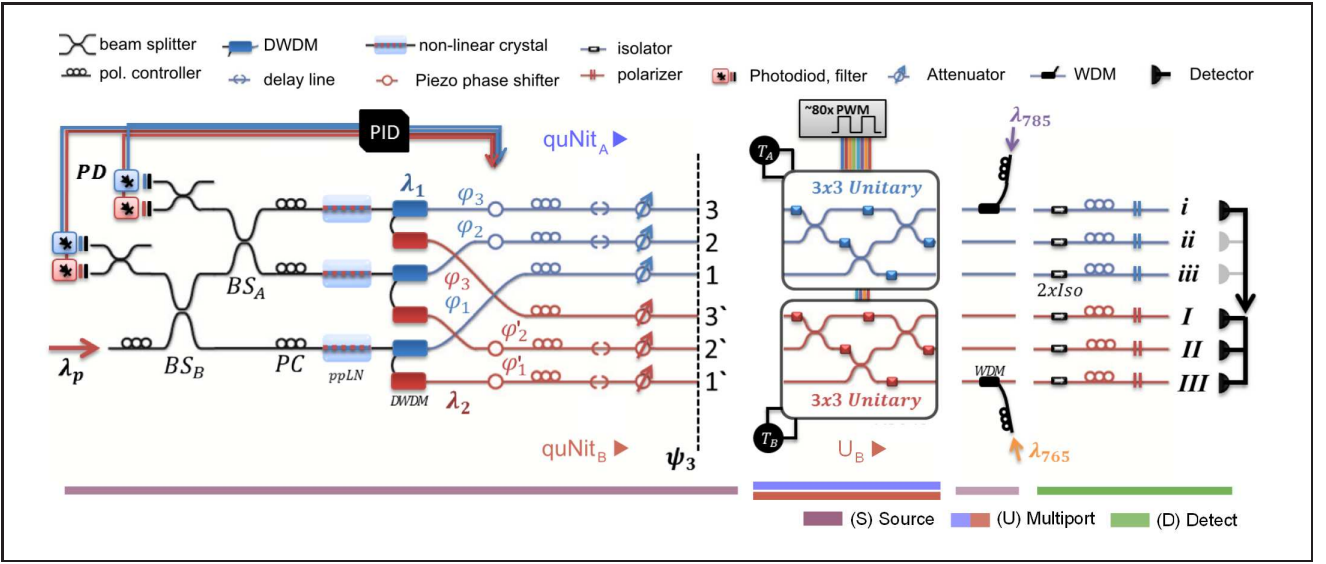}}
\caption{Illustration of the experimental arrangement. All components are directly interfaced via standard optical fiber connectors or directly spliced together for minimal optical loss. (S) A superposition between N photon pair creation events is produced by coherently pumping N non-linear crystals. For each channel the photon pair is separated by wavelength followed by a number of in-fiber components ensuring matching optical properties. Eventually, an entangled two qutrit state is obtained. Using the interference signal of reference light ($\lambda_{765},\lambda_{785}$) inserted backwards into the system a PID controller is actively stabilizing all phases of the in-fiber system. (U) Each qutrit is directly connected to the packaged Multiports performing any arbitrary local unitary transformation onto the system. (D) Eventually, using single photon detectors the measurement concludes when pair detection events are recorded.}
\label{fig:preparation}
\end{figure*}
\subsection{Detection (D)}
The outputs $a=i$ of qutrit A are detected by a free-running, high performance, low noise single photon detector. A detection event will simultaneously trigger three single photon detectors at output modes $b=I,II,III$  of qutrit B opening the detection window for 2.5ns. A detection event at B will be treated as one detected photon pair leading to the count rate $CC_{ab}$ . Output modes $(a,b)=(i,I),(i,II),(i,III)$ are measured simultaneously yielding the corresponding probabilities of joint detection $P_{ab}\propto CC_{ab}$. Generally, $CC_{ab}$ depends on all parameters defining the quantum system. In particular, the relative phases $CC_{ab}=CC_{ab}(\phi_1-\phi_1',\phi_1-\phi_1')$ are of importance. Note, for $a\neq i$ no physical detectors are necessary as explained later.
\section{Preparation and Performance}
The complete setup is illustrated in Figure \ref{fig:preparation}. All free parameters defining the quantum system are under experimental control. The {\bf Source (S)} handles the creation of the entangled state:
\begin{itemize}
\item	The pump power within each crystal controls all amplitudes $A_1,A_2,A_3$. Similar to the splitting ratios the coherence of the pump beam could be controlled by inserting fiber delay lines for other types of quantum states. Here, without exception, we set the amplitudes to an equal superposition state $A_1=A_2=A_3=\frac{1}{\sqrt{3}}$ .
\item	An active PID feedback mechanism ensures complete control and stabilization of all phases $\phi_1,\phi_1',\phi_2,\phi_2',\phi_3,\phi_3'$  of the entangled state (details in \cite{Ref15}).

\end{itemize}
The final state should approximately assume the form 
\begin{equation}
|\psi_N\rangle=\sum_{k=1}^{N}A_ke^{\phi_k-\phi_k'}|kk\rangle\,.
\end{equation}

Two {\bf Multiport (U)} devices are connected to qutrit A and B realizing a local transformations taking the form:
\begin{equation}
U_{A/B}=
\begin{pmatrix}
(a/b)_{11} & (a/b)_{12} & (a/b)_{13} \\
(a/b)_{12} & (a/b)_{22}e^{\phi_{22}} & (a/b)_{23}e^{\phi_{23}} \\
(a/b)_{13} & (a/b)_{23}e^{-\phi_{23}} & (a/b)_{33}e^{\phi_{33}}  
\end{pmatrix}\,,
\end{equation}
with a total of 16 real parameters. Each Multiport can be independently configured to realize different local unitary transformations within $SU(3)$.\\
{\bf Single photon detection (D)} completes the experiment. A unitary transformation $U_{Proj}$  of the detection basis allows for the realization of any projective measurement. Thus, if a certain projective measurement is desired, the Multiport is set to include the appropriate transformation $U_{Proj}$ on top of the transformation of the particular experiment $U\rightarrow U U_{Proj}$ . The existence of noise in the system is dominantly a result of accidental coincidence detection events due to a finite detection time window, where the coincidental clicks do not originate from a produced photon pair. Noise of this type can be quantified and is subtracted for all following measurements and quantities if not stated otherwise. Due to the property of the noise, it can be reduced by decreasing the created pair rates per crystal.\\
In anticipation of future experiments, instead of measuring a particular fixed set of relative phases $CC_{ab}(\phi_1-\phi_1',\phi_1-\phi_1')$ we choose a different approach: all relative phases are scanned stepwise within $0,\cdots,2\pi$ . Thereby, a complete map of the correlation space $CC_{ab}$ is obtained. Afterwards, any particular set of phases can be extracted based on a fit obtained from the complete scan of the space. Any measurement herein will utilize this approach.\\
Before continuing with more complex measurements in the full two-qutrit quantum space and for a better understanding of the system, we present some more straight-forward measurement results:
\begin{itemize}
\item	The previous two-qubit source achieved a state fidelity of $F=(99.44\pm0.06)\%$  \cite{Ref15}. The current qutrit system involves three two-qubit subspaces. Using the Multiport all subspaces A,B,C can be accessed and characterized in a similar way (see \cite{Ref15}). Compared to the theoretically expected perfect state we reach a state fidelity of $F_A=(96.22\pm0.73)\%,\,F_B=(93.33\pm0.88)\%,\,F_C=(96.86\pm0.69)\%$
 
Naturally, the resulting fidelity is lower than the optimized qubit source. The qutrit system has not been optimized for measurement in a qubit subspace. In fact, the remaining part of the unitary space the MP cannot realize are perfect qubit transformations excluding one mode from any interaction. Consequently, any measurement using such transformations will be of less quality when assuming a perfect transformation. We have not further addressed this issue since qubit subspaces will be of little interest building a qutrit system. In any case, a sufficient quality of the state is achieved.
\item Extrapolating the two qubit subspace measurements to the full Hilbert space, one immediately reveals the potential of the Multiport to sustain and manipulate genuine higher dimensional entanglement. With the achieved subspace fidelities and an equal population of every path we can use the results of Ref.\cite{Ref6} in order to show that the Multiport is capable of maintaining a genuine qutrit entangled state. We find a corresponding lower bound on the number of "e-bits", i.e. entangled bits, the universal currency of bipartite entanglement of $E_{oF}(\rho)\geq 1.15$ which is significantly above the maximum obtainable value of an entangled qubit system, as qubits can never encode more than one e-bit. Furthermore reaching the upper bound of qubits would require perfectly prepared pure states and noiseless channels and measurements.
\end{itemize}
\section{Results}
\begin{figure*}[htbp]
\centering
\fbox{\includegraphics[width=\linewidth]{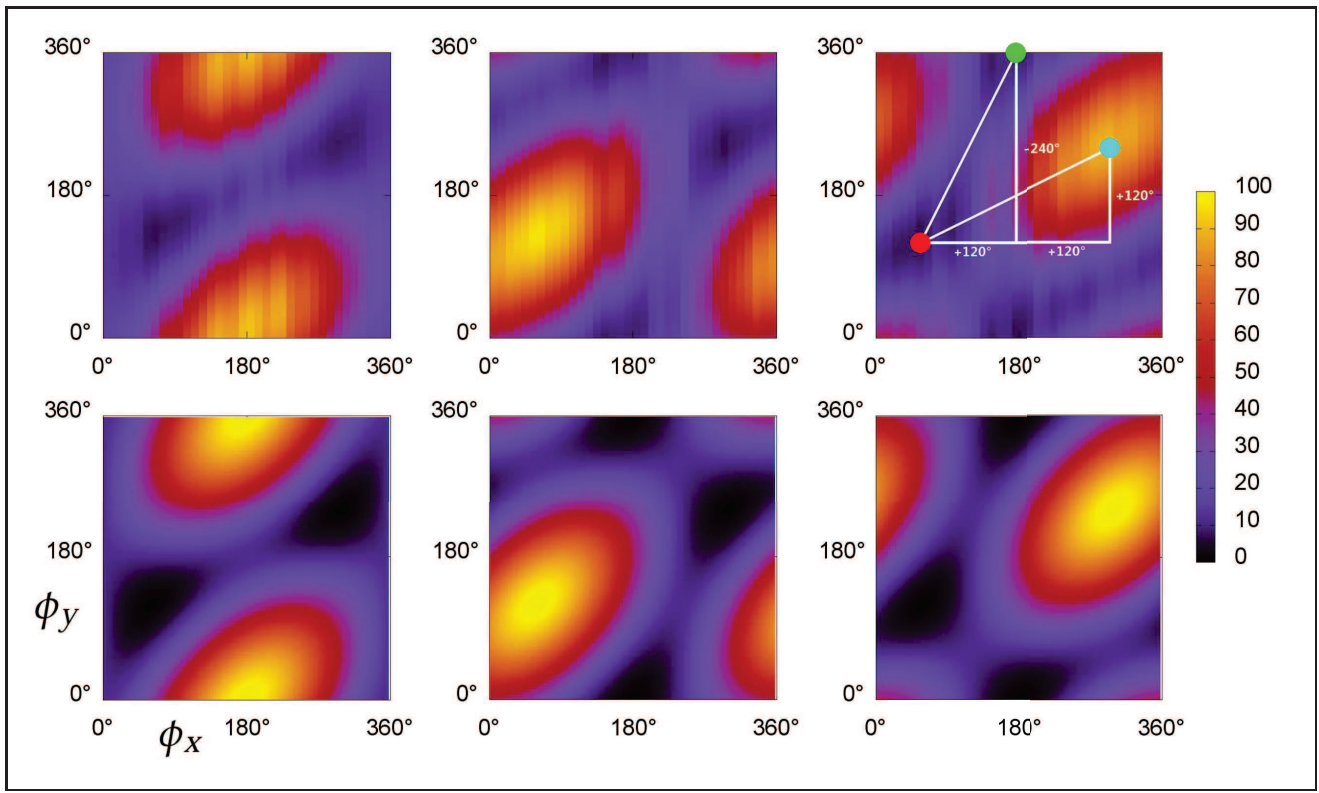}}
\caption{A full scan of the two qutrit correlation space spanned by the two relative phases $CC_{ab}(\phi_x,\phi_y)$ is performed (top row) and compared to theoretic predictions (bottom row). Iteratively, $\phi_x$  is set to $0^{\circ}$,$10^{\circ}$,$20^{\circ}$,$\cdots$,$360^{\circ}$ in steps of $10^{\circ}$. For each value of $\phi_x$, $\phi_y$ is measured in 30 steps between $0$ and $2\pi$. Eventually, a fit $f(\phi_y)$ is applied to the recorded correlation signal $CC_{ab}|^{\phi_y\in(0,2\pi)}_{\phi_x}$. The resulting 36 fitted signals $f(\phi_y)$ are displayed for each $\phi_x$ and for every detector combination $\{(i,I),(i,II),(i,III)\}$ as slice by slice. The maximum coincidence count rates reach approximately $600/8s$ at a $SNR\geq10$. The overlays of the top right picture are used to highlight special characteristics of the correlation space as described in the text. The experiment is calibrated, controlled, stabilized, and performed fully automatically.}
\label{fig:results}
\end{figure*}

For a first demonstration, we perform a complete automated characterization of the higher-order two-qutrit EPR correlations \cite{Ref12}.\\
For its observation a maximally entangled two qutrit state is required along with two 3-dimensional beam splitters and local 3-outcome measurements pointing at the difficulty in observing higher-order correlations.\\
Both Multiports are used to realize the extension of the balanced beam splitter to higher dimensions, the so-called Bellport, taking on the form \cite{Ref12}: 
\begin{equation}
U=\frac{1}{3}
\begin{pmatrix}
1 & 1 & 1 \\
1 & e^{\frac{2\pi}{3}} & e^{\frac{\pi}{3}} \\
1 & e^{\frac{2\pi}{3}} & e^{\frac{2\pi}{3}}  
\end{pmatrix}
\otimes
\begin{pmatrix}
1 & 1 & 1 \\
1 & e^{\frac{2\pi}{3}} & e^{\frac{\pi}{3}} \\
1 & e^{\frac{2\pi}{3}} & e^{\frac{2\pi}{3}}  
\end{pmatrix}
\end{equation}
Note that the matrix elements of the Bellport may be viewed as elements of the discrete Fourier transformation (FT) of the inputs. Thus, if a particle enters one input it will exit the device at any unpredictable output with equal probability (in analogy to the FT of a delta peak). However, in contrast to classical physics, quantum correlations of two entangled particles each entering the similar device allow to impose certain constraints and predictions onto the particle's exit mode. \\
For a detailed analysis, we measure a complete map of the two qutrit correlation space $CC_{ab}(\phi_x,\phi_y)$ spanned by the two relative phases. Iteratively, $\phi_x$  is set to $0^{\circ}$,$10^{\circ}$,$20^{\circ}$,$\cdots$,$360^{\circ}$ in steps of $10^{\circ}$. For each value of $\phi_x$, $\phi_y$ is measured in 30 steps between $0$ and $2\pi$. A fit $f(\phi_y)$ is applied to the recorded correlation signal $CC_{ab}|^{\phi_y\in(0,2\pi)}_{\phi_x}$  while $\phi_x$  is kept constant. Eventually, the resulting 36 fitted signals $f(\phi_y)$ are displayed for each $\phi_x$ and for every detector combination $\{(i,I),(i,II),(i,III)\}$ as slice by slice in Figure \ref{fig:results}. A total of 1080 different measurement points with 8 seconds integration time have been recorded. At maximum a coincidence count rate of around $600/8s$ at a $SNR\geq10$ is achieved. The experiment was calibrated, controlled, stabilized, and performed fully automatically. 
The measurement results (Figure \ref{fig:results}) reveal excellent agreement with theoretic predictions. Notably, given the experimental resolution even the minima are clearly visible. Some particular observations are listed below:
\begin{itemize}
\item	The correlation pattern (Figure \ref{fig:results}) exhibits three distinct correlation maxima at the predicted phase setting (colored dots in Figure \ref{fig:results}) \cite{Ref12}. For example, if detector combination $(i,I)$ exhibits a maximum the other two combinations $(i,II)$ and $(i,III)$ show a minimum in detected coincidences.
\item In principle, the number of possible correlations between detector combinations equals $N!=6$ \cite{Ref12}. However, given the experimental resolution (Figure \ref{fig:results}) we observe three correlation maxima at three distinct phase settings equal to the system's dimension $N=3$. The remaining $N!-N =3$ possible correlations between detectors do not appear, as is expected from quantum theory \cite{Ref12}.
\item	Examining the middle graph of Figure \ref{fig:results}, the maximum correlation appears at $(60^\circ,120^{\circ})$ while minima appear at $(180^\circ,360^\circ)$ and $(300^\circ,240^\circ)$  (as illustrated by the arrows). The relative distance is given by $(0^\circ,0^\circ)$ , $(240^\circ,120^\circ)$ and $(120^\circ,240^\circ)$, as expected.
\item	For a qubit beam splitter, a measurement of the second output is equivalent to a measurement of the first output plus an input phase offset of $180^{\circ}$ \cite{Ref12}, \cite{Ref17}.  Consequently, measuring a certain point $(\phi_x,\phi_y)$ for all possible 9 detector combinations   should be equivalent to measuring the three points  $(\phi_x,\phi_y)$, $(\phi_x+120^\circ,\phi_y+240^\circ)$ and $(\phi_x+240^\circ,\phi_y+120^\circ)$  while recording only joint detection events between $(i,I)$, $(i,II)$ and $(i,III)$. Figure \ref{fig:results} strengthens this assumption which is the reason for the implemented detection scheme using only a single detector for qutrit A.
\item	The observed correlation pattern is similar to the expected classical interference pattern of a three-path Mach-Zehnder-interferometer as previously observed \cite{Ref18}. Here, the pattern is non-classical: it becomes visible solely in coincidence detection. Local observation of single detection events (ideally) does not exhibit any pattern and a photon will be observed with the same probability of 1/3 at any output.
\item	Three path interference results in a higher phase sensitivity $S\propto |dI/d\phi|$ due to  the existence of side lobes \cite{Ref18}. In order to reveal the effect within the two particle correlation pattern the output intensity $I$ is replaced by the number of detected pairs and extracted from the measurement data. The maximal measured phase sensitivity averaged over the three measured detector combinations becomes:
\begin{equation}
\overline{S}=(0.703\pm0.019)rad^{-1}
\end{equation}
The theoretical maximum sensitivity of three-paths interference is given by $S_3^{th}\approx0.782$ . Compared to the qubit case with $S_2^{th}=\frac{1}{2}$ an increase in sensitivity by a factor of $\gamma=1.41\pm0.04$  is achieved.
\end{itemize}
These observations demonstrate the good operation of the Multiport devices and the achievable purity of the input states. The fact that the whole correlation space is accessible also implies that any quantum information protocol, that can be performed using entanglement and local unitary transformations, is directly implementable with our setup.

In addition, we wish to highlight the scalability of the current system towards higher-dimensional quantum systems. The scalability of the source has been previously addressed \cite{Ref15}. While a detailed discussion would be beyond scope of this article, we can provide a representative example of the power of integrated optics in the context of the Multiport design. The integrated Matrix Switch, a device used in the area of communication, redirects light entering one of $N$ inputs to one of $N$ outputs by a combination of tunable MZIs. Matrix Switches up to $32\times32$ in/outputs based on similar Silica waveguide technology have been reported \cite{Ref15e} featuring $2048$ beam splitters, $2048$ phase shifters and $60$cm of total waveguide length integrated on a single chip at an insertion loss below $10$dB. In comparison, the Matrix Switch circuit structure is highly similar to a Multiport. In fact, in terms of device complexity a $32\times32$ Multiport would require half the number of optical devices compared to the above Matrix Switch\footnote{Unfortunately, current-generation Matrix Switches are highly optimized leading to a circuit design not usable as a Multiport. Other designs \cite{Ref15f} have a higher similarity to the Multiport, we will address this issue in a separate article.}. 
An important aspect of interfacing a high-dimensional path-entangled source as ours with multi-port chips is the scaling of the resources for active phase-stabilization. The method we have developed for the N=3 case uses back-illuminating the complete beam path with two lasers of different wavelength injected into the output. The interference signal at 2 of the inputs for both lasers (4 detectors in total, see Fig. 1) is then measured and used for stabilizing the phases of the fibers connecting the source and the multi-port chip. A careful analysis shows that scaling up this design would require only 2 more detectors (one for each wavelength) per additional dimension, but no additional lasers. Thus our active phase-stabilization scheme scales very favourable and contributes to the excellent scalability of our whole approach to higher dimensions.
Additionally, losses of $10$dB are handled by the current source and detection system. Although of relevance for the circuits's design, due to the nature of integrated circuits coherence length and path difference are assumed not to be of major issue. However, a major stepping-stone towards high-dimensional Multiports may be the decreasing process control with increasing circuit sizes and therefore reduced quality of the optical devices. Here, the major concern is the extinction ratio or visibility of the MZIs of the MP. Ultimately, low extinction ratios will reduce the MP's capability to realize arbitrary unitary transformations. In the following we provide an estimate of this effect:
\begin{figure}[htbp]
\centering
\fbox{\includegraphics[width=\linewidth]{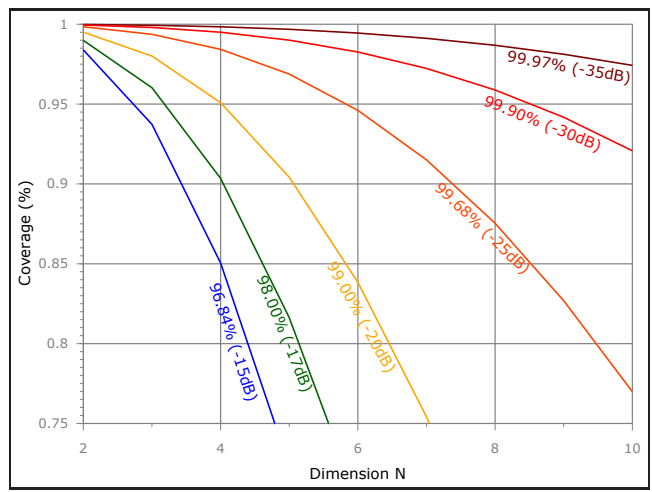}}
\caption{Estimated coverage of the unitary space of an $N\times N$ Multiport of standard design \cite{Ref1,Ref12}. Each plotted line assumes a different extinction ratio for all MZI units inside the Multiport. Phase shifters are assumed to cover the full range of phases as has been experimentally verified. Currently, the characterized Multiport exhibits an average visibility of $\left(96.3\pm 0.01\right)\%$ (extinction ratio of $\left(14.2\pm0.1\right)$dB). However, the current MP implementation has the a first/prototype run. Consequently, we are confident that next-generation devices tailored towards high extinction ratios will provide significant improvements. As an example, a $10$-dimensional Multiport with an extinction ratio of $-30$dB of all MZIs can realize more than $92\%$ of the full unitary $SU(10)$ space.}
\label{fig:scaling}
\end{figure}
The range of unitary operations implementable by the MP can be quantified. We define the "coverage" as the amount of unitary space "covered" by the Multiports:
\begin{align}
cov_N=\frac{\prod_{i=1}^{N^2} \int \limits_{\min{\lambda_i}}^{\max{\lambda_i}}d\lambda_i J_N}{\int_{U(N)}dU_N}\,,
\end{align}
i.e. the ratio between the reachable unitary space given by all parameters and the full unitary space. The factor $J_N$ represents the absolute value of the Jacobi determinant, see \cite{Ref15d} for further details. The limits of $\lambda_i$ are related to the range of reactivities (phases) implementable by each beam splitter (phase shifter). Analyzing a single Multiport we estimate the coverage to $\left(96.3\pm 0.01\right)\%$ of all unitary operations. Here, the limitation stems from the extinction ratios of the Mach-Zehnder interferometers (MZI) while phase shifters cover the full range $0...2\pi$. Using the experimental results we can further extrapolate the coverage of a higher-dimensional Multiport as illustrated in Fig.~(\ref{fig:scaling}). 
Our results thus demonstrate that with experimentally feasible noise levels, moderate scalability of the devices is straightforward. Significant improvements can be expected with further tailoring of next iterations of Multiport chips. Note, however, using a Multiport of high coverage might not even be necessary as long as it can implement the sets of unitaries required for the experiment or quantum information protocol.

\section{Conclusion}
Combining previously developed devices we have successfully implemented a system allowing for an exceptionally high degree of experimental control of an entangled two qutrit, 9-dimensional photonic quantum system. An in-fiber source creates an entangled two qutrit state of any amplitude and phase, while two Multiport devices realize any desired arbitrary local unitary transformation onto the quantum state. We demonstrate the quality of the system by performing a complete characterization of high-order perfect EPR correlations between two qutrits. Results are compared to theoretic prediction. Based on the performance of the current implementation, no direct technical limitations towards higher-dimensional systems become apparent. The source is directly extendible to two ququarts and could be implemented on-chip as well \cite{Ref15}. Compared to other state-of-the-art integrated circuits used in the area of telecommunication \cite{Ref19} higher-dimensional realizations of the Multiport are highly promising for quantum systems of dimension $N$. Further experiments may involve the generation  and  manipulation  of  entangled  states  \cite{Ref20},  state discrimination  \cite{Ref21},  simulation of quantum logic \cite{Ref22}, general optical simulation of spin-1 systems \cite{Ref12}, an optical analog of the Stern-Gerlach experiment \cite{Ref12}, tests of non-contextuality \cite{Ref4}, \cite{Ref12}, realization of general POVM measurements \cite{Ref21}, \cite{Ref24}, or assist in the search for the existence of complete sets of MUBs \cite{Ref25}, \cite{Ref26}. Additionally, the compatibility of the in-fiber source and Multiport to standard fiber networks at the Telecom wavelength will be of further interest for future applications. The design is a viable approach towards higher-dimensional quantum systems and offers great potential towards a higher level of integration and complexity of quantum systems.
\section*{Funding Information}
This work was supported by the ERC (Advanced Grant QIT4QAD, 227844), the Austrian Science Fund FWF within the SFB F40 (FoQuS) and W1210-2 (CoQuS) and through the SIQS grant no. 600645 EU-FP7-ICT. MH acknowledges funding from the Juan de la Cierva fellowship (JCI 2012-14155), the European Commission (STREP "RAQUEL") and the Spanish MINECO Project No. FIS2013-40627-P, the Generalitat de Catalunya CIRIT Project No. 2014 SGR 966. SR is funded by a EU Marie-Curie Fellowship (PIOF-GA-2012- 329851).
\section*{Acknowledgments}
We thank Gerhard Hummer and the Austrian Institute of Technology for their contribution and technical help on the detection system.

\end{document}